\documentclass[]{article}

\usepackage{arxiv}

\usepackage[utf8]{inputenc} % allow utf-8 input
\usepackage[T1]{fontenc}    % use 8-bit T1 fonts
\usepackage{hyperref}       % hyperlinks
\usepackage{url}            % simple URL typesetting
\usepackage{booktabs}       % professional-quality tables
\usepackage{amsfonts}       % blackboard math symbols
\usepackage{nicefrac}       % compact symbols for 1/2, etc.
\usepackage{microtype}      % microtypography
\usepackage{graphicx}
\usepackage{natbib}
\usepackage{doi}
\usepackage{arydshln}
\usepackage{amsmath}
\usepackage{subcaption}
\usepackage{multirow}
\usepackage{rotating}

\title{Revisiting Optimal Allocations for Binary Responses: Insights from Considering Type-I Error Rate Control}

\author{ \href{https://orcid.org/0009-0003-9512-681X}{\includegraphics[scale=0.06]{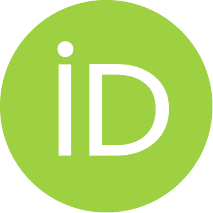}\hspace{1mm}Lukas Pin}\\
	MRC Biostatistics Unit \\
	University of Cambridge\\
	Cambridge, UK \\
	\texttt{lukas.pin@mrc-bsu.cam.ac.uk} \\
	\And
	\href{https://orcid.org/0000-0001-7755-2637}{\includegraphics[scale=0.06]{orcid.pdf}\hspace{1mm}Sofía S. Villar} \\
	MRC Biostatistics Unit \\
	University of Cambridge\\
	Cambridge, UK \\
	\texttt{sofia.villar@mrc-bsu.cam.ac.uk} \\
	\And
    {
    William F. Rosenberger} \\
	Department of Statistics \\
	George Mason University\\
	Fairfax, USA \\
	\texttt{wrosenbe@gmu.edu} \\
}

\renewcommand{\shorttitle}{\textit{Revisiting Optimal Proportions for Binary Responses}}

\hypersetup{
pdftitle={Revisiting Optimal Proportions for Binary Responses},
pdfsubject={biostatistics, RAR},
pdfauthor={Lukas Pin, Sofia Villar, Frank Bretz},
pdfkeywords={Neyman Allocation; RSHIR Allocation; Wald Test; Type-I Error Rate; Patient Benefit},
}

\begin{document}
\maketitle

\begin{abstract}
This work revisits optimal response-adaptive designs from a type-I error rate perspective, highlighting when and how much these allocations exacerbate type-I error rate inflation — an issue previously undocumented. We explore a range of approaches from the literature that can be applied to reduce type-I error rate inflation. However, we found that all of these approaches fail to give a robust solution to the problem. To address this, we derive two optimal allocation proportions, incorporating the more robust score test (instead of the Wald test) with finite sample estimators (instead of the unknown true values) in the formulation of the optimization problem. One proportion optimizes statistical power and the other minimizes the total number failures in a trial while maintaining a fixed variance level. Through simulations based on an early-phase and a confirmatory trial we provide crucial practical insight into how these new optimal proportion designs can offer substantial patient outcomes advantages while controlling type-I error rate. While we focused on binary outcomes, the framework offers valuable insights that naturally extend to other outcome types, multi-armed trials and alternative measures of interest.
\end{abstract}


\section{Introduction}\label{sec1}
While adaptive designs hold promise, their practical application in confirmatory trials lags behind their theoretical prominence \citep[e.g.,][]{Burnett2020, Dimairo2015}.
Response-adaptive randomization (RAR) designs is an example of this phenomenon, having a history spanning nearly a century %, tracing back to the seminal work of 
\citep{Thompson1933} but remaining more a topic of persistent controversy among statisticians \citep{Robertson2023} than a practically chosen design option. This is partly due %the complexity of these designs, which offer a wide variety of approaches for randomizing patients throughout a trial, with each variant influencing final statistical inference differently, and leading 
to the need to balance trade-offs in terms of patient-benefit considerations, power, and type-I error rate as part of achieving acceptability for their practical application. Often discussions of the use of RAR have focused on the trade-off between patient-benefit metrics and statistical power while the type-I error rate has been ignored in the theoretical literature. From a regulatory perspective, type-I error rate control is an absolute minimal requirement for any trial design, without which concerns about power or patient benefit become tangential.

In this paper, we provide new insights for clinical trial designers by proposing two novel optimal RAR designs that explicitly consider type-I error rate control alongside power and patient benefits.
Our focus is two-armed trials with binary endpoints, which are commonly used in clinical research. 
In a literature review on PubMed, we identified over 2,200 papers that discuss Phase III trials with a binary endpoint published in the last 10 years. 
%We reexamine from a new perspective the paper by \cite{Rosenberger2001}.
%In our literature review on PubMed, we identified over 2,200 papers on that discuss Phase III trials with a binary endpoint published in the last 10 years only, see Appendix \ref{sec:litreview}.
%\citep{rombach2020binary}, especially in the context of confirmatory studies. 
We reexamine from a new perspective the paper by \cite{Rosenberger2001}, who first proposed an approach based on targeting optimal allocation proportions, and is one of the most cited papers (335) in the RAR literature.

In Section \ref{sec:Optimal}, we present two optimal allocation proportions that were presented in \cite{Rosenberger2001} and serve as the foundation of our analysis. The first, Neyman allocation, maximizes the statistical power of a widely used Wald test. The second, RSHIR allocation, an abbreviation of the names of authors of \cite{Rosenberger2001}, minimizes the expected number of patient failures during the trial while ensuring the test maintains a predefined variance level. Both allocations are derived for the Wald test. Other have reported issues with its use as an inferential method following response-adaptive trials \citep{antognini2018wald}. Here, we reveal a novel finding: using Neyman or RSHIR allocation further exacerbates type-I error rate inflation to a degree that may surprise readers. To our knowledge, this issue has not been documented in the prior literature.

Section \ref{sec:Corrections} illustrates the type-I error rate performance of approaches that have originally been suggested for related issues, such as non-zero variance estimation. %These include the\say{Agresti \& Caffo} correction in Section \ref{sec:Agresti}, enforcing non-zero variance estimators in Section \ref{sec:nonzeroVariance}, increasing the burn-in period in Section \ref{sec:burnin}, or using an alternative test statistic in Section \ref{sec:AlternativeTest}. 
However, none of these approaches lead to satisfactory type-I error rate control.
To address these concerns, in Section \ref{sec:AlternativeOptimalRAR} we explore how to derive optimal proportions for an alternative test statistic and demonstrate how the principles behind the two optimal proportions can be adapted to mitigate statistical issues. In Section \ref{sec:TrialExample}, we provide practical guidance for using optimal proportions in designing some clinical trial examples, highlighting their potential advantages and limitations. Finally, Section \ref{sec:discuss} concludes with a discussion of the implications of our findings for both the design and evaluation of RAR-based trials.

Our main contribution is the development of a procedure and test statistic that preserves the type-I error rate for binary response RAR designs. 
Note that the structure of this paper follows the way in which we arrived at this procedure.
%template of how we think about solving the problem: first, identify the problem; second, check existing solutions; third, derive a new solution if existing solutions are inadequate; fourth, verify and implement the new solution. 
We emphasize, that in practice, one should first ensure that type-I error rates are controlled before power and patient-benefit gains are considered.

\section{Notation and Framework for Optimal Response Adaptive Randomization}\label{sec:Optimal}

In this and the next two sections, we consider a two-armed binary trial with a fixed sample size $n = n_0 + n_1$, where $n_0$ and $n_1$ are the sample sizes for the control and treatment arms, respectively. The responses of individual patients are modeled as random variables $Y_{ki}$. Here, $k \in \{0,1\}$ indicates the treatment ($k=1$) or control ($k=0$), and $i \in \{1, \dots, n\}$ indexes the patients. For each arm $k \in \{0,1\}$, the random variables $Y_{k1},...,Y_{kN}$ are assumed to be independent and identically distributed (i.i.d.) as Bernoulli random variables
\begin{equation*}\label{Eq:RARIID}
    Y_{ki} \mathop{\sim}\limits^{\mathrm{i.i.d.}} \text{Bern}(p_k), \quad i = 1, \dots, n_k; \quad k = 0,1; \quad n_0 + n_1 = n,
\end{equation*}
where $p_k$ represents the success probability of arm $k$, and $q_k = 1 - p_k$ is the probability of failure.

Out of the $2 \times n$ potential outcomes (corresponding to two treatments and \( n \) patients), only \( n \) are observed in the trial, since for each patient we can observe the outcome under only one treatment. The treatment assignment for patient $i$ is indicated by a binary variable $A_{i}$, where $A_{i} = 1$ if the patient is assigned to the experimental arm and $A_{i} = 0$ otherwise. The probability of assigning a patient to treatment $k$ is therefore $ P(A_{i} = 1)$, and $P(A_{i} = 0 ) = 1 - P(A_{i} = 1)$. Because randomization  
%Randomization is a cornerstone of clinical trials, as it 
minimizes bias, reduces confounding, and ensures comparability between treatment groups, we focus here only on designs that satisfy:  %Any fully randomized design, including RAR, must satisfy 
$P(A_{i} = k ) > 0$ for all $i=1,\dots,n$ and $k=0,1$. %, ensuring every treatment has a nonzero probability of being assigned. 
The simplest (non-adaptive) randomization scheme is complete randomization (CR), where $P(A_{i} = k ) = 1/2$ for all $i$ and $k$ in a two-arm trial. 
%This corresponds to assigning treatments through a fair coin toss. CR is widely used for its simplicity and because it is maintaining equipoise at the trial's outset. 
CR attains an equal allocation proportion (on average) and it is often assumed to maximize statistical power \citep{Berger2021, Friedman2015} and control the type-I error rate.

In contrast, RAR dynamically adjusts allocation probabilities based on accumulating trial data. We assume patients enter the trial sequentially, with outcomes observed immediately, to explore the full potential of RAR designs.
%, but this can be relaxed to patients arriving in groups and delayed outcomes. 
Let $a^{(j)} = \{a_1, \dots, a_j\}$ and $y^{(j)} = \{y_1, \dots, y_j\}$ represent the allocation and response histories after $j$ patients, respectively, with $a^{(0)}$ and $y^{(0)}$ as empty sets. Under RAR, the allocation probability for patient $i$ depends on prior allocations and responses $P(A_{i} = 1 | a^{(i-1)}, y^{(i-1)})$
for $i = 2, \dots, n$. For at least the first patient, the probabilities are typically set to $P(A_{1} = 1) = 1/2$, reflecting initial equipoise. However, prior knowledge from earlier studies can inform alternative initial probabilities.
%Moreover, one could consider employing a burn-in period for the first $B \in \{1,\cdots,n\}$ patients to stabilize the MLEs before adapting the probability distribution to decrease the probability of sampling in the wrong direction, see Section \ref{sec:burnin}.

At the trial's conclusion, a statistical test is chosen for final inference. The two-sample Wald test is widely used in RAR literature for evaluating optimal designs \citep{Hu2006}.
%, Pin_Deming2024
This test considers a measure of interest based on the success probabilities, denoted as $f(p_0, p_1)$. In this paper, we focus (as does \cite{Rosenberger2001}) on the mean parameter difference $f(p_0, p_1) = p_1 - p_0$, though other measures such as the log relative risk or log odds ratio can also be used as suggested by \cite{Rosenberger2001}. A complete table of optimal allocation proportions is given by \citet[p.~325]{Pin_Deming2024}.
The general null hypothesis is 
\begin{equation*}
    H_0: f(p_0, p_1) = f_0 \quad \text{versus} \quad H_1: f(p_0, p_1) \neq f_0,
\end{equation*}
which can be evaluated using the test statistic
\begin{equation}\label{eq:GeneralWald}
    Z = \frac{f(\hat{p}_0, \hat{p}_1) - f_0}{\sqrt{\widehat{\mathrm{Var}}(f(\hat{p}_0, \hat{p}_1))}},
\end{equation}
where $\hat{p}_k$ are the MLEs of $p_k$, and $\widehat{\mathrm{Var}}(f(\hat{p}_0, \hat{p}_1))$ is the estimated variance of the test statistic. For the mean difference $f(p_0, p_1) = p_1 - p_0$ and $f_0=0$, the test statistic \eqref{eq:GeneralWald} simplifies to
\begin{equation*}\label{eq:ZtestBern}
    Z_1 = \frac{\hat{p}_1 - \hat{p}_0}{\sqrt{\frac{\hat{p}_0 \hat{q}_0}{n_0} + \frac{\hat{p}_1 \hat{q}_1}{n_1}}}.
\end{equation*}

In practice, RAR is typically combined with early stopping and allows for the allocation of patients in groups and delayed outcomes. However, to fully explore the potential of these algorithms, we will use the framework described above. In the next section, we present two optimal sampling proportions for the test statistic \( Z_1 \), as derived by \cite{Tschuprow1923, Neyman1934, Robbins1952} and \cite{Rosenberger2001}.

\subsection{Two Optimal Proportions for the Wald Test}\label{sec:OptimalProportions}
%\subsection{Neyman Allocation}\label{sec:DerivationNeyman1}

The first allocation we discuss is Neyman allocation \citep{Neyman1934, Robbins1952}. The aim is to fix the asymptotic variance of the the treatment difference estimator (or equivalently the inverse of the Fisher information, as in Jennison and Turnbull's formulation \citeyearpar[p. 327]{Jennison1999}), 
%maximize the statistical power of the Wald test $Z_1$, 
%i.e. the probability to reject the null hypothesis when $p_0 \neq p_1$. The aim is %Maximizing power of the test is equivalent to minimizing the variance of the test statistic 
given by $s_{T}^{2} = (p_0 q_0) / n_0 + (p_1 q_1)/n_1$
\citep[pp. 910]{Rosenberger2001}, and under that constraint optimize 
%Hence, the optimization problem can be formulated as
\begin{equation*}
    \text{min}_{\rho} n_{0}+n_{1}  \textbf{ \quad s.t. \quad } s_{T}^{2}\leq C,
\end{equation*}
where $(1-\rho) = n_{0} / n$, $\rho  = n_{1} /n $ and $C \in \mathbb{R}^{+}$.
%, which can be rewritten as
%\begin{align}\label{eq:varianceRHO}
%    \frac{\rho\sigma_{0}^{2}+(1-\rho)\sigma_{1}^{2}}{\rho(1-\rho)C} &= n.
%\end{align}
Deriving with respect to 
%$\rho_{N_{1}}$ 
$\rho$ yields
\begin{align*}
  %  \frac{(p_0q_0-p_1q_1)\rho(1-\rho) - (\rho p_0q_0 + (1-\rho)p_1q_1)(1-2 \rho)C}{\rho^2(1-\rho)^2C^2} = 0 \nonumber \\
   % \frac{(1-\rho)^2}{\rho^2}  =  \frac{p_0q_0}{p_1q_1} \nonumber \\
    \rho_{N_{1}} = \frac{\sqrt{p_1q_1}}{\sqrt{p_0q_0}+\sqrt{p_1q_1}}, \label{eq:NeymanWald}
\end{align*}
Neyman allocation for the Wald type test $Z_1$.

%\subsection{RSHIR Allocation}\label{sec:RSHIRWALD}

The second optimal allocation we introduce is RSHIR allocation \citep{Rosenberger2001}, which minimizes the expected number of failures in a trial with binary responses, subject to a constraint on the variance of the Wald test. This allocation is determined by solving the following optimization problem, ensuring the variance remains below a pre-specified threshold \( C \): 
\[
\text{min}_{\rho} \, n_0(1 - p_0) + n_1(1 - p_1) \quad \text{subject to} \quad s_T^2 \leq C,
\]
where \( n_0 = (1 - \rho) \cdot n \), \( n_1 = \rho \cdot n \), and \( C \in \mathbb{R}^+ \). 
%After rewriting the variance constraint in terms of \( \rho \) 
%to Equation \eqref{eq:varianceRHO} 
%the objective function to minimize can then be expressed as:
%\[
%\text{min}_{\rho} \, \frac{\left((1 - \rho)(1 - p_0) + \rho(1 - p_1)\right)\left(\rho p_0(1 - p_0) + (1 - \rho) p_1(1 - p_1)\right)}{\rho (1 - \rho) C}.
%\]
%To find the optimal allocation proportion \( \rho \), we differentiate the objective function with respect to \( \rho \) and set the derivative to zero
%\[
%- \frac{(1 - p_0) p_1 (1 - p_1)}{\rho^2 C} + \frac{(1 - p_0) p_0 (1 - p_1)}{(1 - \rho)^2 C} = 0
%\]
%which simplifies to:
%\[
%\frac{p_1}{p_0} = \frac{\rho^2}{(1 - \rho)^2}.
%\]
%Driving with respect to and then solving for \( \rho \), 
Solving the optimization problem leads to the RSHIR allocation for the Wald test $Z_1$
%optimal allocation is given by
\begin{equation*}\label{eq:RSHIRWald}
    \rho_{R_{1}} = \frac{\sqrt{p_1}}{\sqrt{p_0} + \sqrt{p_1}}.
\end{equation*}
%RSHIR allocation for the Wald test $Z_1$.
Notice that to derive the optimal proportions, we utilized the theoretical variance of the test statistic $Z_1$ and the true success probabilities. However, these parameters are typically unknown in a real trial setting. %Section \ref{sec:Targeting} discusses how to estimate and target these proportions, while Section \ref{sec:AlternativeOptimalRAR} focuses on deriving optimal proportions based on estimators needed the calculation of the test statistic, specifically its variance.

\subsection{Targeting Procedures}\label{sec:Targeting}

If the true parameters \(p_0\) and \(p_1\) were known, they could be directly substituted into the optimal allocation proportions $\rho_{N_{1}}$ and $\rho_{R_{1}}$, to guide the sampling process throughout the trial. However, since these parameters are typically unknown, they must be estimated. 

One common approach is to use MLEs to obtain parameter estimates \(\hat{p}_0\) and \(\hat{p}_1\). In the \textit{sequential maximum likelihood procedure (SMLE)}, the parameters are re-estimated sequentially after observing the outcomes of each patient. Before assigning a patient \(j \in \{B, \ldots, n-1\}\), the current estimates \(\hat{p}_0(j)\) and \(\hat{p}_1(j)\) are plugged into the optimal allocation proportion formula to compute an updated sampling proportion, denoted as \(\hat{\rho}_{N_1}(j)\) or \(\hat{\rho}_{R_1}(j)\). This proportion is then used to allocate the next patient $j+1$. 

Two more recently developed methods are the \textit{doubly adaptive biased coin design (DBCD)}, introduced by \cite{Eisele1994, Eisele1995} and modified by \cite{Hu2004};
%. 
%This approach leverages a distance metric to measure the discrepancy between the current sampling proportion \(\hat{\rho}(j)\) and the actual patient allocation \(n(j) = (n_{0}(j), n_{1}(j))\), which reflects the allocations after observing the outcomes of the first \(j\) patients. By adjusting the sampling proportion to reduce this discrepancy, the DBCD ensures faster convergence towards the theoretical optimal allocation.
%We utilize 
and the \textit{efficient randomized-adaptive design (ERADE)} introduced by \cite{Hu2009}. ERADE is the discretized adaptation of the doubly adaptive biased coin design (DBCD) by \cite{Eisele1994} aiming to decrease the variance of the design. Given a parameter $\alpha \in (0,1)$, the probability of assigning patient $j+1$ to treatment 1 is determined as follows:
\begin{equation*} 
    \rho(n_{1}(j),\hat{\rho}(j)) = 
    \begin{cases}
			\alpha \hat{\rho}(j), & \text{if $n_{1}(j)/j >  \hat{\rho}(j)$,}  \\
            \hat{\rho}(j), & \text{if $n_{1}(j)/j =  \hat{\rho}(j)$,} \\
            1-\alpha (1-\hat{\rho}(j)), & \text{if $n_{1}(j)/j <  \hat{\rho}(j)$,}
		 \end{cases}
\end{equation*}
where $\hat{\rho}(j)$ is an estimated optimal proportion and $n_{1}(j)$ the number of patients allocated to arm 1 after the allocation of $j$ patients.
Following the recommendation of \cite{Burman1996}, the parameter $\alpha$ is typically chosen within the range of $0.4$ to $0.7$. We choose $\alpha = 0.5$.

%In the next section, we will use the \textit{efficient randomized-adaptive design (ERADE)} proposed by \cite{Hu2009}. ERADE is a discretized version of Hu and Zhang's DBCD aiming to decrease the variance of the design. 
Both DBCD and ERADE effectively drive the design towards the theoretical optimal allocation. 
%, even when incorporating a burn-in period. 
These targeting procedures enable faster convergence to the desired allocation proportions compared to simpler methods. For a more detailed discussion of these designs, we refer interested readers to \cite{Rosenberger2016}.

\subsection{Simulation Results}

In this section, we present simulation results for the Wald test under three allocation rules: CR (\(\rho_{CR}\)), Neyman (\(\rho_{N_{1}}\)), and RSHIR (\(\rho_{R_{1}}\)), using ERADE as the targeting procedure. A significance level of \(\alpha = 0.05\) was used for the Wald test and will be used for all future tables and figures in this paper. The simulations in this section were conducted with a minimal burn-in period of \(B = 4\), two patients per arm allocated using permuted block randomization, and each scenario was simulated 10,000 times, ensuring a Monte Carlo error of less than 0.5\% for both power and type-I error rate \citep{Morris2019}. We use a burn-in of 2 patients per arm consistently throughout the paper. There is no consensus in the literature on the optimal burn-in size for response-adaptive trials \citep{Tang2025}. We intentionally use the minimal burn-in (2 patients per arm) to investigate the maximal extent of type-I error rate inflation.

Table \ref{tab:WaldSimulation} presents the results for type-I error rate across the entire parametric space. Additionally, it reports results for a fixed $p_0 = 0.2$ ($p_0 = 0.7$) under an increasing (decreasing) sequence of treatment effects to illustrate differences in power.
\begin{table}[htp!]
    \centering
    \small
    \renewcommand{\arraystretch}{1.5} 
    \setlength{\tabcolsep}{6pt}   
    \begin{tabular}{ccccccccccc}
        \hline
        \multicolumn{11}{c}{Testing with $Z_1$} \\
        \hline
         $p_0$ & $p_1$ & \multicolumn{3}{c}{Type-I Error Rate or Power} & \multicolumn{3}{c}{$n_1/n$} & \multicolumn{3}{c}{ENS} \\
         \hline
         & & $\rho_{CR}$ & $\rho_{N_{1}}$ & $\rho_{R_{1}}$ & $\rho_{CR}$ & $\rho_{N_{1}}$ & $\rho_{R_{1}}$ & $\rho_{CR}$ & $\rho_{N_{1}}$ & $\rho_{R_{1}}$ \\
         \hline
        \textbf{0.1} & \textbf{0.1} & \textbf{5.0\%} & \textbf{68.2\%} & \textbf{68.1\%} & 0.5 (0) & 0.47 (0.1477) & 0.46 (0.1471) & 5 & 5 & 5 \\ 
        \textbf{0.2} & \textbf{0.2} & \textbf{5.9\%} & \textbf{82.2\%} & \textbf{80.0\%} & 0.5 (0) & 0.46 (0.1570) & 0.48 (0.1525) & 10 & 10 & 10 \\ 
        \textbf{0.3} & \textbf{0.3} & \textbf{6.3\%} & \textbf{72.0\%} & \textbf{66.8\%} & 0.5 (0) & 0.47 (0.1432) & 0.48 (0.1336) & 15 & 15 & 15  \\ 
        \textbf{0.4} & \textbf{0.4} & \textbf{6.2\%} & \textbf{64.7\%} & \textbf{53.0\%} & 0.5 (0) & 0.47 (0.132) & 0.48 (0.1063) & 20 & 20 & 20 \\ 
        \textbf{0.5} & \textbf{0.5} & \textbf{6.4\%} & \textbf{61.9\%} & \textbf{38.6\%} & 0.5 (0) & 0.47 (0.1261) & 0.49 (0.0757) & 25 & 25 & 25 \\ 
        \textbf{0.6} & \textbf{0.6} & \textbf{6.0\%} & \textbf{65.0\%} & \textbf{26.6\%} & 0.5 (0) & 0.47 (0.1326) & 0.49 (0.0490) & 30 & 30 &  30 \\ 
        \textbf{0.7} & \textbf{0.7} & \textbf{6.1\%} & \textbf{71.9\%} & \textbf{17.8\%} & 0.5 (0) & 0.47 (0.1438) & 0.49 (0.0295) & 35 & 35 & 35 \\ 
        \textbf{0.8} & \textbf{0.8} & \textbf{6.1\%} & \textbf{82.1\%} & \textbf{10.6\%} & 0.5 (0) & 0.47 (0.1567) & 0.49 (0.0127) & 40 & 40 & 40 \\ 
        \textbf{0.9} & \textbf{0.9} & \textbf{4.8\%} & \textbf{68.3\%} & \textbf{5.1\%} & 0.5 (0) & 0.47 (0.1477) & 0.49 (0.0033) & 45 & 45 & 45 \\ 
         \hdashline 
         0.2 & 0.1 & 17.5\% & 84\% & 82.9\% & 0.5 (0) & 0.33 (0.1308) & 0.32 (0.1308) & 7.5 & 8.3 & 8.4 \\ 
         0.2 & 0.3 & 15.3\% & 80\% & 76.7\% & 0.5 (0) & 0.55 (0.1513) & 0.57 (0.1420) & 12.5 & 12.7 & 12.9 \\ 
         0.2 & 0.5 & 65.4\% & 88.3\% & 84.9\% & 0.5 (0) & 0.62 (0.1342) & 0.71 (0.0907)& 17.5 & 19.3 & 20.7 \\
         0.2 & 0.7 & 97.1\% & 98.3\% & 97.8\% & 0.5 (0) & 0.56 (0.1486) & 0.79 (0.0476) & 22.5 & 23.9 & 29.8 \\
         0.7 & 0.2 & 97.2\%& 98.5\%& 97.9\% & 0.5 (0) & 0.38 (0.1408) & 0.19 (0.0408) & 22.5 & 25.5 &  30.2 \\ 
         0.7 & 0.4 & 62.1\% & 85.9\% & 72.0\% & 0.5 (0) & 0.52 (0.1381) & 0.33 (0.0507) & 27.5 & 27.2 & 30.0 \\ 
         0.7 & 0.6 & 13.9\% & 70.4\% & 26.6\% & 0.5 (0) & 0.51 (0.1394) & 0.45 (0.0372)  & 32.5 & 32.5 &  32.8 \\
         0.7 & 0.8 & 15.0\% & 78.9\% & 22.0\% & 0.5 (0) & 0.38 (0.1396) & 0.52 (0.0213) & 37.5 & 36.9 &  37.6 \\
         \hline
    \end{tabular}
    \caption{Power or Type-I Error Rate, proportion allocated to the treatment arm $n_1/n$, and expected number of successes (ENS) for $n=50$, a burn-in size of $2$ patients per arm and different settings of $p_0$ and $p_1$.}
    \label{tab:WaldSimulation}
\end{table}
We observe that both allocation strategies achieve their intended objectives. Using Neyman allocation results in the highest power, whereas RSHIR allocation maximizes the expected number of successes. However, it is important to note that the outcomes of these procedures exhibit high variability, particularly when Neyman allocation is applied.

\begin{figure}[ht]
    \centering
    % First subfigure
    %\begin{subfigure}[b]{0.45\textwidth}
    \begin{subfigure}[b]{0.85\textwidth}
        \centering
        \includegraphics[width=\textwidth]{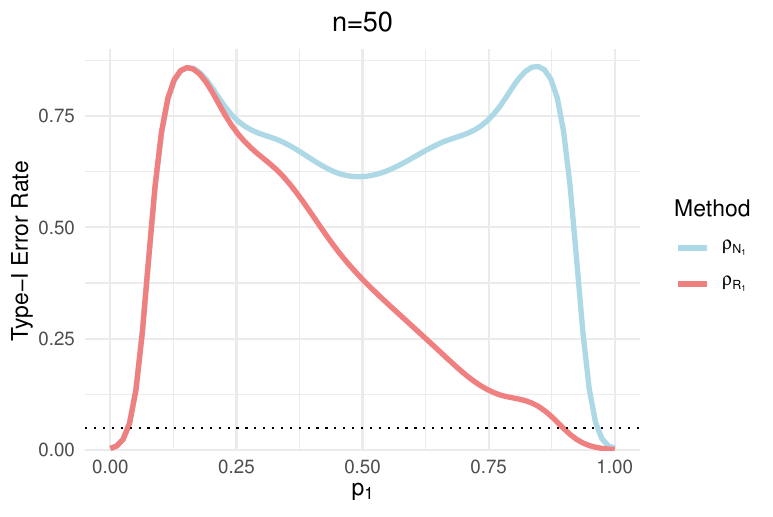}
        %\caption{Type-I error rates for $\rho_{N_{1}}$ and $\rho_{R_{1}}$ testing with the Wald test.}
        \label{fig:typeIerror}
    \end{subfigure}
    \hfill % Add horizontal space between subfigures
    % Second subfigure
    %\begin{subfigure}[b]{0.45\textwidth}
    \begin{subfigure}[b]{0.85\textwidth}
        \centering
        \includegraphics[width=\textwidth]{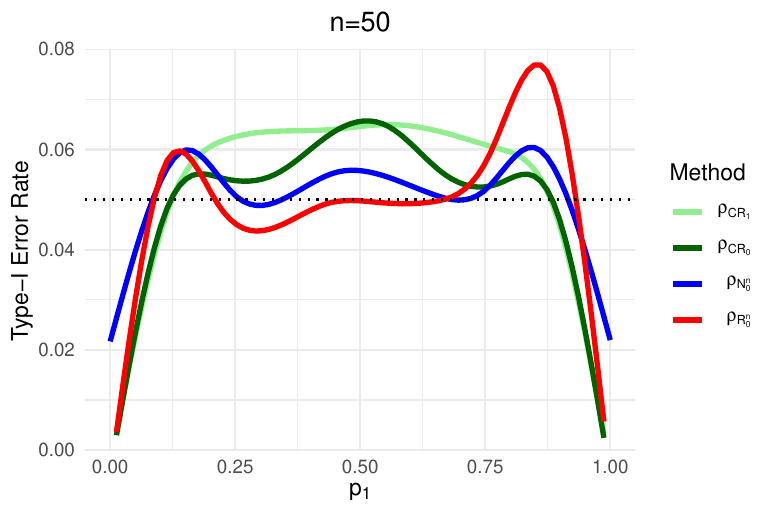}
       % \caption{Type-I error rates for CR testing with the Wald test ($\rho_{CR_{1}}$) and CR ($\rho_{CR_{0}}$), $\rho^{n}_{R_0}$ and $\rho^{n}_{N_0}$ testing with the score test.}
        \label{fig:typeIerror_Z0}
    \end{subfigure}
    \caption{Comparison of Type-I error rates of: the Neyman proportion testing with the Wald test $\rho_{N_{1}}$, the RSHIR proportion testing with the Wald test $\rho_{N_{1}}$, complete randomization testing with the Wald test $\rho_{CR_1}$, complete randomization testing with the score test $\rho_{CR_0}$, the Neyman-like proportion testing with the score test $\rho^{n}_{N_{0}}$ and the RSHIR-like proportion testing with the score test $\rho^{n}_{R_{0}}$. In the both figures a burn-in size of $2$ patients per arm was used.}
    \label{fig:TypeIFigures}
\end{figure}

As noted by \citet[page 30]{fleiss1981}, the Wald test $Z_1$ can lead to type-I error rate inflation even when CR is used. Despite a Monte Carlo error of 0.5\%, we observe type-I error rate inflation in both scenarios of \(p_0 = p_1 = 0.2\) and \(p_0 = p_1 = 0.7\), with an inflation exceeding 1\% in the latter case. Furthermore, the use of optimal allocation exacerbates type-I error rate inflation to a level that is indeed surprising. Specifically, Neyman allocation \(\rho_{N_{1}}\) inflates the type-I error rate to over 81\% in the case of \(p_0 = p_1 = 0.2\), but the inflation with RSHIR allocation is almost equally high. The type-I error rate inflation across the whole parametric space is shown in Figure \ref{fig:TypeIFigures}. 
To the best of our knowledge, this issue has not been previously documented and renders the use of these procedures in their current form impractical.

%In the next Section we present known approaches that deal with the type-I error rate inflation of the Wald test $Z_1$ and discuss their interplay with the derived optimal allocations $\rho_{N_{1}}$ and $\rho_{R_{1}}$.

\section{Evaluating Existing Approaches to Reduce Type-I Error Rate Inflation}\label{sec:Corrections}

In this section, we present existing approaches that can help reduce the inflation of the type-I error rate. Although these methods have not been explicitly proposed for type-I error rate control, their application seems closely related, as they address issues that contribute to type-I error rate inflation, such as non-zero variance estimators or reduced adaptivity.

%To address the inflated type-I error rate of the Wald test, one could think to use the \say{Agresti \& Caffo} correction, see Section \ref{sec:Agresti}, which ensures non-zero estimators of the variances during testing, or to sample with equal probability while one of the variance estimators is equal to zero, see Section \ref{sec:nonzeroVariance}. As a third option increasing the burn-in size $B$ can help stabilize the estimators and promote a shift toward equal allocation, as discussed in Section \ref{sec:burnin}. Lastly, one could use the score instead of the Wald test for testing, see Section \ref{sec:AlternativeTest}. 

\subsection{Agresti \& Caffo Correction}\label{sec:Agresti}

\cite{Agresti2000} 
%propose adding pseudo-observations to each arm to improve the coverage of standard confidence intervals and ensure non-zero variance estimators. This method 
use the same test statistic $Z_1$ but modify the original MLE variance estimator by incorporating one artificial failure and one artificial success for each arm, adjusting the success probability estimates to \(\hat{p}_i^a = (s_i + 1)/(n_i + 2)\), where \(s_i\) is the number of successes. This adjustment ensures that the estimated success probabilities fall within the open interval \((0,1)\) and that therefore the estimated variances \(\hat{p}_i^a(1 - \hat{p}_i^a)\) are non-zero.

However, the Agresti \& Caffo estimator, \(\hat{p}_i^a\), only converges to the MLE asymptotically. For small sample sizes, especially with success probabilities near 0 or 1, the Agresti \& Caffo estimator \(\hat{p}_i^a\) may introduce considerable bias and lead to incorrect test decisions. As shown in Table \ref{tab:WaldSimulationAdapted}, while the correction reduces type-I error rate inflation, it does not control the type-I error rate at a significance level of \(\alpha = 0.05\).

\begin{table}[htp!]
    \centering
    \small
    \renewcommand{\arraystretch}{1.5} 
    \setlength{\tabcolsep}{6pt}   
    \begin{tabular}{cccccccccccccc}
         \hline
         $p_0$ & $p_1$ & \multicolumn{6}{c}{Type-I Error Rate or Power} & \multicolumn{6}{c}{ENS} \\
         \cline{3-14}
         & & \multicolumn{2}{c}{Agresti \& Caffo} & \multicolumn{2}{c}{$B=12$} & \multicolumn{2}{c}{$B=30$} & \multicolumn{2}{c}{Agresti \& Caffo} & \multicolumn{2}{c}{$B=12$} & \multicolumn{2}{c}{$B=30$} \\
         \cline{3-14}
         & & $\rho_{N_{1}}$ & $\rho_{R_{1}}$ & $\rho_{N_{1}}$ & $\rho_{R_{1}}$ & $\rho_{N_{1}}$ & $\rho_{R_{1}}$ & $\rho_{N_{1}}$ & $\rho_{R_{1}}$ & $\rho_{N_{1}}$ & $\rho_{R_{1}}$ & $\rho_{N_{1}}$ & $\rho_{R_{1}}$ \\
         \hline
        \textbf{0.1} & \textbf{0.1} & \textbf{0.3\%} & \textbf{0.1\%} & \textbf{51.7\%} & \textbf{52.5\%} & \textbf{19.0\%} & \textbf{19.0\%} & 5 & 5 & 5 & 5 & 5 & 5 \\
         \textbf{0.2} & \textbf{0.2} & \textbf{2.8\%} & \textbf{0.4\%} & \textbf{45.3\%} & \textbf{44.3\%} & \textbf{12.2\%} & \textbf{12.1\%} & 10 & 10 & 10 & 10 & 10 & 10 \\ 
        \textbf{0.3} & \textbf{0.3} & \textbf{6.1\%} & \textbf{1.3\%} & \textbf{25.5\%} & \textbf{24.5\%} & \textbf{8.7\%} & \textbf{9.0\%} & 15 & 15 & 15 & 15 & 15 & 15 \\
        \textbf{0.4} & \textbf{0.4} & \textbf{6.6\%} & \textbf{3.0\%} & \textbf{14.3\%} & \textbf{14.3\%} & \textbf{7.8\%} & \textbf{7.7\%} & 20 & 20 & 20 & 20 & 20 & 20 \\
        \textbf{0.5} & \textbf{0.5} & \textbf{7.1\%} & \textbf{5.3\%} & \textbf{11.1\%} & \textbf{9.3\%} & \textbf{6.9\%} & \textbf{7.1\%} & 25 & 25 & 25 & 25 & 25 & 25 \\
        \textbf{0.6} & \textbf{0.6} & \textbf{6.9\%} & \textbf{7.5\%} & \textbf{14.8\%} & \textbf{6.7\%} & \textbf{7.7\%} & \textbf{6.6\%} & 30 & 30 & 30 & 30 & 30 & 30 \\
        \textbf{0.7} & \textbf{0.7} & \textbf{6.1\%} & \textbf{11.5\%} & \textbf{25.8\%} & \textbf{6.0\%} & \textbf{8.9\%} & \textbf{6.2\%} & 35 & 35 & 35 & 35 & 35 & 35 \\
        \textbf{0.8} & \textbf{0.8} & \textbf{3.1\%} & \textbf{9.1\%} & \textbf{43.9\%} & \textbf{5.7\%} & \textbf{11.8\%} & \textbf{5.4\%} & 40 & 40 & 40 & 40 & 40 & 40 \\
        \textbf{0.9} & \textbf{0.9} & \textbf{0.3\%} & \textbf{3.4\%} & \textbf{51.7\%} & \textbf{4.8\%} & \textbf{19.4\%} & \textbf{4.4\%} & 45 & 45 & 45 & 45 & 45 & 45 \\
         \hdashline 
         0.2 & 0.1 & 1.4\% & 0.5\% & 0.4\% & 60.8\% & 29.6\% & 28.5\% & 8.4 & 8.4 & 8.2 & 8.2 & 7.9 & 7.9 \\ 
         0.2 & 0.3 & 6.9\% & 2.3\% & 40.4\% & 41.1\% & 19.0\% & 20.1\% & 12.8 & 12.8 & 12.8 & 12.8 & 12.7 & 12.7 \\ 
         0.2 & 0.5 & 36.2\% & 25.1\% & 72.2\% & 71.3\% & 67.6\% & 67.6\% & 19.2 & 20.7 & 19.2 & 19.8 & 18.3 & 18.9 \\
         0.2 & 0.7 & 81.5\% & 75.8\% & 98.0\% & 97.4\% & 97.5\% & 97.4\% & 23.8 & 29.8 & 24.2 & 27.4 & 23.2 & 25.5 \\
         %0.7 & 0.2 & 82.2\% & 80.8\% & 97.6\% & 97.2\% & 97.7\% & 97.4\% & 25.4 & 30.2 & 24.6 & 27.8 & 23.7 & 25.7 \\ 
         %0.7 & 0.4 & 47.5\% & 66.9\% & 66.2\% & 62.6\% & 63.5\% & 61.5\% & 27.3 & 30 & 27 & 28.9 & 27.4 & 28.6 \\ 
         %0.7 & 0.6 & 11.7\% & 55.3\% & 25.3\% & 13.0\% & 15.7\% & 12.9\% & 32.5 & 32.8 & 32.3 & 32.6 & 32.4 & 32.6 \\
         %0.7 & 0.8 & 6.1\% & 47.6\% & 41.9\% & 15.1\% & 19.6\% & 14.7\% & 36.9 & 37.6 & 37.1 & 37.6 & 37.3 & 37.5 \\
         \hline
    \end{tabular}

    \vspace{1cm}

       \begin{tabular}{cccccccccccccc}
         \hline
         $p_0$ & $p_1$ & \multicolumn{6}{c}{Type-I Error Rate or Power} & \multicolumn{6}{c}{ENS} \\
         \cline{3-14}
         & & \multicolumn{2}{c}{Agresti \& Caffo} & \multicolumn{2}{c}{$B=12$} & \multicolumn{2}{c}{$B=120$} & \multicolumn{2}{c}{Agresti \& Caffo} & \multicolumn{2}{c}{$B=12$} & \multicolumn{2}{c}{$B=120$} \\
         \cline{3-14}
         & & $\rho_{N_{1}}$ & $\rho_{R_{1}}$ & $\rho_{N_{1}}$ & $\rho_{R_{1}}$ & $\rho_{N_{1}}$ & $\rho_{R_{1}}$ & $\rho_{N_{1}}$ & $\rho_{R_{1}}$ & $\rho_{N_{1}}$ & $\rho_{R_{1}}$ & $\rho_{N_{1}}$ & $\rho_{R_{1}}$ \\
         \hline
        \textbf{0.1} & \textbf{0.1} & \textbf{0.6\%} & \textbf{0.4\%} & \textbf{76.5\%} & \textbf{76.8\%} & \textbf{7.9\%} & \textbf{8.2\%} & 20 & 20 & 20 & 20 & 20 & 20 \\
        \textbf{0.2} & \textbf{0.2} & \textbf{3.7\%} & \textbf{1.2\%} & \textbf{44.8\%} & \textbf{44.3\%} & \textbf{5.9\%} & \textbf{6.5\%} & 40 & 40 & 40 & 40  & 40  & 40 \\ 
        \textbf{0.3} & \textbf{0.3} & \textbf{8.4\%} & \textbf{2.3\%} & \textbf{23.2\%} & \textbf{24.2\%} & \textbf{5.7\%} & \textbf{5.0\%} & 60 & 60 & 60 & 60 & 60 & 60 \\
        \textbf{0.4} & \textbf{0.4} & \textbf{7.0\%} & \textbf{3.5\%} & \textbf{12.9\%} & \textbf{12.4\%} & \textbf{5.3\%} & \textbf{5.7\%} & 80 & 80 & 80 & 80 & 80 & 80 \\
        \textbf{0.5} & \textbf{0.5} & \textbf{7.1\%} & \textbf{5.0\%} & \textbf{9.7\%} & \textbf{7.6\%} & \textbf{5.5\%} & \textbf{5.4\%} & 100 & 100 & 100 & 100 & 100 & 100 \\
        \textbf{0.6} & \textbf{0.6} & \textbf{7.2\%} & \textbf{8.3\%} & \textbf{13.6\%} & \textbf{5.7\%} & \textbf{5.7\%} & \textbf{5.4\%} & 120 & 120 & 120 & 120 & 120 & 120 \\
        \textbf{0.7} & \textbf{0.7} & \textbf{8.0\%} & \textbf{15.1\%} & \textbf{24.4\%} & \textbf{5.1\%} & \textbf{5.5\%} & \textbf{5.3\%} & 140 & 140 & 140 & 140 & 140  & 140  \\
        \textbf{0.8} & \textbf{0.8} & \textbf{3.7\%} & \textbf{10.2\%} & \textbf{44.2\%} & \textbf{5.1\%} & \textbf{6.1\%} & \textbf{5.4\%} & 160 & 160 & 160 & 160 & 160 & 160 \\
        \textbf{0.9} & \textbf{0.9} & \textbf{0.8\%} & \textbf{6.2\%} & \textbf{76.2\%} & \textbf{5.1\%} & \textbf{7.7\%} & \textbf{5.3\%} & 180 & 180 & 180 & 180 & 180 & 180 \\
         \hdashline 
         0.2 & 0.1 & 6.8\% & 5.2\% & 80.5\% & 80.3\% & 54.7\% & 53.9\% & 33.9  & 34.1 & 33.6 & 33.7 & 31.5 & 31.7 \\ 
         0.2 & 0.3 & 14.1\% & 10.3\% & 56.3\% & 56.7\% & 38.5\% & 39.8\% & 51.2  & 51.9 & 51.8 & 51.9  & 50.6 & 51.0 \\ 
         0.2 & 0.5 & 49.6\% & 42.6\% & 99.5\% & 99.4\% & 99.5\% & 99.6\% & 78.1   & 84.3 & 78.4 & 81.5   & 73.3  & 76.4 \\
         0.2 & 0.7 & 89.7\% & 87.1\% & 100.0\%  & 100.0\% & 100.0\% & 100.0\% & 97.0   & 121.9 & 98.6 & 112.7 & 93.3 & 102.7 \\
         %0.7 & 0.2 & 92.1\% & 92.1\% & 100.0\%  & 100.0\% & 100.0\% & 100.0\% & 102.9 & 122.8 & 99.6 & 113.3 & 93.8 & 102.9  \\ 
         %0.7 & 0.4 & 67.4\% & 91.5\% & 99.4\% & 99.2\% & 99.3\% & 99.3\% & 108.5 & 120.1 & 107.7 & 115.3  & 109.2  & 114.3 \\ 
         %0.7 & 0.6 & 19.2\% & 37.0\% & 42.4\% & 32.7\% & 32.7\% & 31.7\% & 129.7   & 131.0 & 129.3 & 130.5 & 129.7 & 130.3 \\
         %0.7 & 0.8 & 11.8\% & 41.6\% & 56.7\% & 38.1\% & 39.5\% & 37.2\% & 147.3  & 150.6 & 148.1 & 150.2 & 149.3 & 150.3  \\
         \hline
    \end{tabular}

    \caption{Comparison of type-I error rate or power and ENS for various settings of $p_0$ and $p_1$: The tables provide results for the Agresti \& Caffo correction and different burn-in periods ($B=12$, $B=30$, $B=120$) for two sample sizes, $n=50$ (upper table) and $n=200$ (lower table). For the Agresti \& Caffo correction a minimal burn of size $2$ patients per arm was used.}
    \label{tab:WaldSimulationAdapted}
\end{table}

\subsection{Ensuring Existence of Variance Estimators}\label{sec:nonzeroVariance}

\cite{Rosenberger2001} also suggest sampling with equal probability when one of the variance estimators is zero, indicating that all observations in that sample are identical. %In clinical trials with binary endpoints, this situation arises when an arm consists entirely of non-responders (all zeros) or responders (all ones). 
This often happens in small samples or at the beginning of a larger trial, especially with success rates near zero or one. Under these conditions, the adaptive design effectively simplifies to CR.  %due to equal probability sampling between arms.

\begin{table}[htp!]
    \centering
    \small
    \renewcommand{\arraystretch}{1.5} 
    \setlength{\tabcolsep}{6pt}   
    \begin{tabular}{ccccccccccc}
        \hline
        \multicolumn{11}{c}{Testing with $Z_1$} \\
        \hline
         $p_0$ & $p_1$ & \multicolumn{3}{c}{Type-I Error Rate or Power} & \multicolumn{3}{c}{$n_1/n$} & \multicolumn{3}{c}{ENS} \\
         \hline
         & & $\rho_{CR}$ & $\rho_{N_{1}}$ & $\rho_{R_{1}}$ & $\rho_{CR}$ & $\rho_{N_{1}}$ & $\rho_{R_{1}}$ & $\rho_{CR}$ & $\rho_{N_{1}}$ & $\rho_{R_{1}}$ \\
         \hline
        \textbf{0.1} & \textbf{0.1} & \textbf{5.0\%} & \textbf{4.8\%} & \textbf{4.0\%} & 0.5 (0) & 0.49 (0.0036) & 0.49 (0.0044) & 5 & 5 & 5 \\ 
        \textbf{0.2} & \textbf{0.2} & \textbf{5.9\%} & \textbf{8.2\%} & \textbf{8.1\%} & 0.5 (0) & 0.49 (0.0043) & 0.49 (0.0062) & 10 & 10 & 10 \\ 
        \textbf{0.3} & \textbf{0.3} & \textbf{6.3\%} & \textbf{8.6\%} & \textbf{9.0\%} & 0.5 (0) & 0.49 (0.0023) & 0.49 (0.0.005) & 15 & 15 & 15 \\ 
        \textbf{0.4} & \textbf{0.4} & \textbf{6.2\%} & \textbf{7.2\%} & \textbf{8.2\%} & 0.5 (0) & 0.49 (0.0008) & 0.49 (0.0032) &  20 & 20 & 20 \\ 
        \textbf{0.5} & \textbf{0.5} & \textbf{6.4\%} & \textbf{7.1\%} & \textbf{7.4\%} & 0.5 (0) & 0.49 (0.0004) & 0.49 (0.0020) & 25 &25 &25  \\ 
        \textbf{0.6} & \textbf{0.6} & \textbf{6.0\%} & \textbf{7.7\%} & \textbf{6.2\%} & 0.5 (0) &  0.49 (0.0007) & 0.49 (0.0013) & 30 & 30 & 30 \\
        \textbf{0.7} & \textbf{0.7} & \textbf{6.1\%} & \textbf{8.6\%} & \textbf{6.1\%} & 0.5 (0) & 0.49 (0.0022) & 0.49 (0.0008) & 35 & 35 & 35 \\ 
        \textbf{0.8} & \textbf{0.8} & \textbf{6.1\%} & \textbf{7.9\%} & \textbf{6.3\%} & 0.5 (0) & 0.49 (0.0042) & 0.49 (0.0005) & 40 & 40 & 40 \\ 
        \textbf{0.9} & \textbf{0.9} & \textbf{4.8\%} & \textbf{4.7\%} & \textbf{5.0\%} & 0.5 (0) & 0.49 (0.0036) & 0.49 (0.0003) & 45 & 45 & 45 \\ 
         \hdashline 
         0.2 & 0.1 & 17.5\% & 18.5\% & 17.7\% & 0.5 (0) & 0.44 (0.0036) & 0.44 (0.0049) & 7.5 & 7.8 & 7.8 \\ 
         0.2 & 0.3 & 15.3\% & 18.1\% & 18.4\% & 0.5 (0) & 0.53 (0.0032) & 0.54 (0.0054) & 12.5 & 12.6 & 12.7 \\ 
         0.2 & 0.5 & 65.4\% & 67.6\% & 68.5\% & 0.5 (0) & 0.55 (0.0022) & 0.61 (0.0036)& 17.5 & 18.2 & 19.1 \\
         0.2 & 0.7 & 97.1\% & 97.3\% & 97.4\% & 0.5 (0) & 0.53 (0.0032) & 0.64 (0.0029) & 22.5 & 23.2 & 26.0 \\
     %    0.7 & 0.2 & 97.2\%& 97.8\%& 97.2\% & 0.5 (0) & 0.45 (0.0032) & 0.34 (0.0028) & 22.5 & 23.7 &  26.5 \\ 
    %     0.7 & 0.4 & 62.1\% & 63.5\% & 62.7\% & 0.5 (0) & 0.51 (0.0016) & 0.41 (0.0022) & 27.5 & 27.4 & 28.8 \\ 
    %     0.7 & 0.6 & 13.9\% & 16.0\% & 12.9\% & 0.5 (0) & 0.51 (0.0016) & 0.47 (0.0011)  & 32.5 & 32.4 &  32.6 \\
    %     0.7 & 0.8 & 15.0\% & 18.6\% & 14.9\% & 0.5 (0) & 0.45 (0.0031) & 0.51 (0.0007) & 37.5 & 37.3 &  37.6 \\
         \hline
    \end{tabular}
    \caption{Results for sampling with equal probability if one of the two variance estimators is 0. Power or Type-I Error Rate, proportion allocated to the treatment arm $n_1/n$, and expected number of successes (ENS) for $n=50$, a burn-in size of $2$ patients per arm and different settings of $p_0$ and $p_1$.}
    \label{tab:WaldSimulation_CorrectedVariance}
\end{table}
Table \ref{tab:WaldSimulation_CorrectedVariance} shows that adopting equal-probability sampling reduces type-I error rate inflation from 80\% to 8\% when success rates are not at extremes. However, this adjustment brings power and ENS metrics Neyman and RSHIR allocation closer to CR. 
%Neyman and RSHIR allocations provide minimal gains in power or patient benefit.

Hence, for smaller samples, this adjustment effectively attenuates type-I error rate inflation by reducing the adaptive design to CR and (even for larger samples). However, as shown in Table \ref{tab:WaldSimulation_CorrectedVariance} it does not fully eliminate type-I error rate inflation and it considerably reduces any advantages from an adaptive approach.

\subsection{Burn-In Periods to Stabilize Estimators and Decrease Adaptivity}\label{sec:burnin}

A burn-in period is defined as the initial allocation of the first $B \in \{1,\dots,n\}$ patients with equal probability to each arm. To ensure an equal number of patients ($B/2$) in each arm, we enforce balance using methods such as simple random allocation or permuted block randomization. This step stabilizes the MLEs before adapting the allocation probabilities to reduce the likelihood of assigning patients in the wrong direction. The minimal burn-in size for RSHIR is one patient per arm ($B = 2$), whereas for Neyman, it is two patients per arm ($B = 4$), see Table \ref{tab:WaldSimulation}. %\cite{Rosenberger2001} recommend using six patients per arm ($B = 12$).
% and extending the burn-in period further as long as one of the two sample variances remains zero. 
Table \ref{tab:WaldSimulationAdapted} demonstrates that (as the other approaches) the procedures with using larger burn-in periods still result in inflated type-I error rates. Moreover, increasing the burn-in size %helps reduce type-I error rate inflation but 
comes at the cost of diminishing patient-benefit gains (RSHIR) or power gains (Neyman). For Table \ref{tab:WaldSimulationAdapted} we used a burn-in period of six patients per arm ($B = 12$), as recommend by \cite{Rosenberger2001}, and 60\% of the total trial size to demonstrate the the asymptotic behavior. 
In smaller trials, this trade-off often forces a near complete convergence to a CR design in an attempt to control the type-I error rate, though full control may not always be achievable, see Section \ref{sec:AlternativeTest}. In larger trials, the same burn-in size allows for adaptive sampling while retaining error control, see Table \ref{tab:WaldSimulationAdapted}. 
%In the limit (\(B = n\)), the algorithm reduces to CR, but type-I error rate inflation may still occur when testing with \(Z_1\) \citep[page 30]{fleiss1981}.

\subsection{Using the Score Test: Testing with a Pooled Variance Estimator}\label{sec:AlternativeTest}

%In \cite{Rosenberger2001}, patient allocation proportions are derived based on the variance of the Wald-type test statistic $Z_1$, as originally suggested by \cite{Neyman1934}. However, 
\cite{Eberhardt1977} and \citet[page 30]{fleiss1981} observed that the Wald-type test statistic $Z_1$ often exhibits an inflated type-I error rate in non-adaptive settings. To mitigate this issue, they propose testing the null hypothesis $H_0: p_0 = p_1$ using the score test statistic \citep[p. 14]{Agresti2002}, which is defined as
\begin{equation*}\label{eq:Ztest0Bern}
    Z_0 = \frac{\hat{p}_1 - \hat{p}_0}{\sqrt{ \hat{p} \hat{q} \left(\frac{1}{n_0} + \frac{1}{n_1}\right)}},
\end{equation*}
where $\hat{p}$ is the overall proportion of successes in the trial, calculated as the total number of successes divided by the trial size $n$, and $\hat{q} = 1 - \hat{p}$. Comparing the CR results from Table~\ref{tab:WaldSimulation} with those from Table~\ref{tab:1_for_ZO_Simulation} and Figure~\ref{fig:TypeIFigures} confirms that this test provides superior control over the type-I error rate.

The score test $Z_0$, is commonly used \citep{Agresti2002} and it can be shown that the squared statistic, $Z_0^2$, is equivalent to the widely used Pearson chi-squared test \citep[p.~119]{scott2020statistics}.
In contrast, the variance of the test statistic $Z_1$ that was used by \cite{Tschuprow1923, Robbins1952, Rosenberger2001} incorporates the individual estimates of $p_0$ and $p_1$ in the estimator of the variance of the test statistic, enabling the construction of confidence intervals that align with the test decision. 
%Specifically, if the null hypothesis is not rejected, the confidence interval will include $p_1 - p_0 = 0$. 
While using the Wald test typically achieves higher power compared to the score test, it does so at the expense of inflating the type-I error rate, even in non-adaptive settings \citep[page 30]{fleiss1981}.

\cite{Rosenberger2001} suggest using the allocation proportion $\rho_{N_1}$ with the $Z_0$ test. However, combining $\rho_{N_1}$ with the score test does not effectively control the type-I error rate or maximize power, as shown in Table \ref{tab:1_for_ZO_Simulation}. This is because $\rho_{N_1}$ is optimized for $Z_1$, not $Z_0$. Similarly, the RSHIR allocation proportion $\rho_{R_1}$, which is derived using variance of the Wald test, also fails with $Z_0$.

\begin{sidewaystable}
    \centering
    \small
    \renewcommand{\arraystretch}{1.5} 
    \setlength{\tabcolsep}{5pt}   
    \begin{tabular}{ccccccccccccccccc}
        \hline
        \multicolumn{17}{c}{Testing with $Z_0$} \\
        \hline
         $p_0$ & $p_1$ & 
         \multicolumn{5}{c}{Type-I Error Rate or Power} & 
         \multicolumn{5}{c}{$n_1/n$} & 
         \multicolumn{5}{c}{ENS} \\
         \hline
         & & 
         $\rho_{CR}$ & $\rho_{N_{1}}$ & $\rho_{R_{1}}$ & $\rho_{N^{n}_{0}}$ & $\rho_{R^{n}_{0}}$ & 
         $\rho_{CR}$ & $\rho_{N_{1}}$ & $\rho_{R_{1}}$ & $\rho_{N^{n}_{0}}$ & $\rho_{R^{n}_{0}}$ & 
         $\rho_{CR}$ & $\rho_{N_{1}}$ & $\rho_{R_{1}}$ & $\rho_{N^{n}_{0}}$ & $\rho_{R^{n}_{0}}$ \\
         \hline
        \textbf{0.1} & \textbf{0.1} & 4.6\% & 0.3\% & 0.1\% & 6.8\% & 6.0\% & 0.5 (0) & 0.47 (0.1477) & 0.46 (0.1471) & 0.49 (0.0047) & 0.49 (0.0036) & 5 & 5 & 5 & 5 & 5 \\
        \textbf{0.2} & \textbf{0.2} & 5.5\% & 2.9\% & 0.4\% & 5.0\% & 5.2\% & 0.5 (0) & 0.46 (0.1568) & 0.48 (0.1525) & 0.49 (0.0033) & 0.48 (0.1525) & 10 & 10 & 10 & 10 & 10 \\
        \textbf{0.3} & \textbf{0.3} & 5.6\% & 6.7\% & 1.2\% & 4.9\% & 4.7\% & 0.5 (0) & 0.47 (0.1432) & 0.48 (0.1336) & 0.49 (0.0015) & 0.49 (0.0005) & 15 & 15 & 15 & 15 & 15 \\
        \textbf{0.4} & \textbf{0.4} & 6.2\% & 7.0\% & 2.9\% & 5.3\% & 4.8\% & 0.5 (0) & 0.47 (0.132) & 0.48 (0.1063) & 0.49 (0.0006) & 0.49 (0.0005) & 20 & 20 & 20 & 20 & 20 \\
        \textbf{0.5} & \textbf{0.5} & 6.0\% & 6.7\% & 5.4\% & 5.5\% & 4.6\% & 0.5 (0) & 0.47 (0.1261) & 0.49 (0.0757) & 0.49 (0.0003) & 0.49 (0.0017) & 25 & 25 & 25 & 25 & 25 \\
        \textbf{0.6} & \textbf{0.6} & 5.5\% & 7.2\% & 8.6\% & 5.3\% & 5.0\% & 0.5 (0) & 0.47 (0.1326) & 0.49 (0.0490) & 0.49 (0.0006) & 0.49 (0.0040) & 30 & 30 & 30 & 30 & 30 \\
        \textbf{0.7} & \textbf{0.7} & 5.6\% & 6.7\% & 13.2\% & 4.8\% & 5.5\% & 0.5 (0) & 0.47 (0.1437) & 0.49 (0.0293) & 0.49 (0.0015) & 0.49 (0.0074) & 35 & 35 & 35 & 35 & 35 \\
        \textbf{0.8} & \textbf{0.8} & 5.6\% & 3.2\% & 10.3\% & 5.3\% & 6.8\% & 0.5 (0) & 0.47 (0.1567) & 0.49 (0.0127) & 0.49 (0.0033) & 0.48 (0.0126) & 40 & 40 & 40 & 40 & 40 \\
        \textbf{0.9} & \textbf{0.9} & 5.1\% & 0.3\% & 5.5\% & 6.1\% & 7.9\% & 0.5 (0) & 0.47 (0.1477) & 0.49 (0.0033) & 0.49 (0.0047) & 0.47 (0.0145) & 45 & 45 & 45 & 45 & 45 \\
         \hdashline
         0.2 & 0.1 & 17.2\% & 1.5\% & 0.5\% & 17.4\% & 18.6\% & 0.5 (0) & 0.32 (0.1324) & 0.32 (0.1296) & 0.54 (0.0041) & 0.53 (0.0028) & 7.5 & 8.4 & 8.4 & 7.3 & 7.4 \\
         0.2 & 0.3 & 13.9\% & 7.0\% & 2.1\% & 12.5\% & 12.1\% & 0.5 (0) & 0.56 (0.1488) & 0.58 (0.1391) & 0.46 (0.0023) & 0.47 (0.0011) & 12.6 & 12.8 & 12.9 & 12.3 & 12.4 \\
         0.2 & 0.5 & 64.1\% & 35.4\% & 24.7\% & 60.7\% & 60.0\% & 0.5 (0) & 0.62 (0.1337) & 0.71 (0.0913) & 0.43 (0.0016) & 0.50 (0.0020) & 17.5 & 19.4 & 20.8 & 16.5 & 17.6 \\
         0.2 & 0.7 & 97.0\% & 84.2\% & 80.3\% & 96.4\% & 96.2\% & 0.5 (0) & 0.55 (0.1494) & 0.79 (0.0472) & 0.46 (0.0023) & 0.60 (0.0050) & 22.5 & 23.8 & 29.8 & 21.5 & 25.1 \\
         \hline
    \end{tabular}
    \caption{Comparison of five testing procedures using the score test: Type-I Error Rate or Power, proportion allocated to the treatment arm $n_1/n$, and expected number of successes (ENS) for $n=50$ and a burn-in size of $2$ per arm.}
    \label{tab:1_for_ZO_Simulation}
\end{sidewaystable}

\section{Revisiting Optimal RAR Using an Error Control Approach}\label{sec:AlternativeOptimalRAR}

Implementation of optimal allocation proportions requires estimating the unknown parameters $p_0$ and $p_1$. These parameters are estimated using maximum likelihood estimation. Using DBCD or ERADE the resulting procedure converges to the target theoretical optimal allocation $\rho$, e.g. $\rho_{N_{1}}$ or $\rho_{R_{1}}$. 

The RAR procedures first introduced in this section substitutes the MLEs $\hat{p}_0$ and $\hat{p}_1$ for the unknown parameters $p_0$ and $p_1$ in the optimization problems from Section \ref{sec:OptimalProportions}
%\ref{sec:DerivationNeyman1} and \ref{sec:RSHIRWALD} 
directly, finding the optimal proportion at each step of the trial, which we denote as $\rho^{n}$ since the proportion depends on the sample size $n$. Hence, we take a finite sample approach based on the estimators rather than an asymptotic approach based on the unknown true values.

Moreover, we will use the variance of the score test $Z_0$ rather than $Z_1$.
Since the test statistic of the score test $Z_0$ has an estimated variance that depends on the estimator $\hat{p}$ of unknown value of $p$ under the null hypothesis, we leverage this estimated variance as the basis for defining our optimization problem.
This approach enables us to optimize the behavior of the score test for departures from the null hypothesis while controlling type-I error rate. 
%Unlike the standard optimization approach, which is expressed as a function of the unknown parameters in the nonlinear case and selects suboptimal RAR procedures to converge to a solution, our method directly defines the optimization problem as a function of the MLEs rather than the unknown parameters.

\subsection{Neyman-like Allocation}

In Section \ref{sec:AlternativeTest} we demonstrated that we cannot simply use $\rho_{N_{1}}$ as an allocation rule to maximize power and test with the score test $Z_0$ to control the type-I error rate.
To address this limitation, we propose deriving a Neyman-like allocation tailored for the $Z_0$ test, minimizing its variance. 
First, we note that the estimator $\hat{p}$ is simply the total number of successes divided by the trial size $n$, but can be expressed using the weighted average of the MLEs of each treatment arm $\hat{p} := (n_0 \hat{p}_0 + n_1 \hat{p}_1)/n$.
We can then rewrite the squared variance of the test statistic as follows
\begin{align}
    \frac{n \hat{p} \hat{q}}{n_0 n_1} &= \frac{n \left(\frac{n_0 \hat{p}_0 + n_1 \hat{p}_1}{n}\right) \left(1 - \frac{n_0 \hat{p}_0 + n_1 \hat{p}_1}{n}\right)}{n_0 n_1} \nonumber \\
    &= \frac{\hat{p}_0}{\rho^{n} n} + \frac{\hat{p}_1}{(1 - \rho^{n}) n} - \frac{(1 - \rho^{n}) \hat{p}_0^2}{\rho^{n} n} - \frac{2 \hat{p}_0 \hat{p}_1}{n} - \frac{\rho^{n} \hat{p}_1^2}{(1 - \rho^{n}) n}. \label{eq:VARZ0RHO}
\end{align}
We then derive the allocation proportion $\rho^{n}$ by minimizing the variance with respect to $\rho^{n}$ and solving for it, which yields
\begin{equation*}
    \rho^{n}_{N_0} = \frac{\sqrt{\hat{p}_0 \hat{q}_0}}{\sqrt{\hat{p}_0 \hat{q}_0} + \sqrt{\hat{p}_1 \hat{q}_1}}.
\end{equation*}
Interestingly, it follows that $\rho^{n}_{N_0} = 1 - \hat{\rho}_{N_1}$ where $\hat{\rho}_{N_1}$ is the optimal solution form Section \ref{sec:Optimal} with plugged-in estimators $\hat{p}_0$ and $\hat{p}_1$.

\subsection{RSHIR-like Allocation}\label{sec:newRSHIR}

%Using the proportion $\rho_{N_1}$ and testing with $Z_1$ also leads to substantial type-I error rate inflation, as demonstrated in Section \ref{sec:Optimal}. 
In this section we derive a proportion that minimizes the number of expected failures given a fixed variance constraint when testing with $Z_0$.

The second optimization problem from Section \ref{sec:OptimalProportions} remains unchanged but the variance in the constraint is now the variance of test statistic $Z_0$, specifically we can use it as formulated in equation \eqref{eq:VARZ0RHO} and rewrite the optimization problem to
\begin{equation*}
    \text{min}_{\rho^{n}} \, \frac{\left((1 - \rho^{n})(1 - \hat{p}_0) + \rho^{n}(1 - \hat{p}_1)\right)\left( \frac{\hat{p}_0}{\rho^{n}} + \frac{\hat{p}_1}{(1-\rho^{n})} - \frac{(1-\rho^{n})\hat{p}_0^2}{\rho^{n}} - 2\hat{p}_0\hat{p}_1 - \frac{\rho^{n} \hat{p}_1^2}{(1-\rho^{n})}\right)}{C}.
\end{equation*}

deriving with respect to $\rho^{n}$ yields
\[
0 = \frac{1}{C} \Bigg[ (\hat{p}_0 - \hat{p}_1) \left( \frac{\hat{p}_0 (1 - \hat{p}_0 + \rho^{n} \hat{p}_0)}{\rho^{n}} + \frac{\hat{p}_1 - \rho^{n} \hat{p}_1^2}{1-\rho^{n}} - 2\hat{p}_0\hat{p}_1 \right) 
\]
\[
+ (1 - \hat{p}_0 + \rho^{n} \hat{p}_0 - \rho^{n} \hat{p}_1) \left( \frac{\hat{p}_1(1-\hat{p}_1)}{(1-\rho^{n})^2} - \frac{\hat{p}_0(1-\hat{p}_0)}{(\rho^{n})^2} \right) \Bigg].
\]
Even though we cannot find a close form solution for $\rho^{n}$ we can solve for it computationally and obtain $\rho^{n}_{R_0}$ (see Supplementary Material).
In Figure \ref{fig:AllProportions} we demonstrate how the proportions differ theoretically and present simulation results in the next Section. 
\begin{figure}[ht!]
    \centering
    % First figure
    %\begin{subfigure}[t]{0.32\textwidth}
    \begin{subfigure}[t]{0.45\textwidth}
        \centering
        \includegraphics[width=\textwidth]{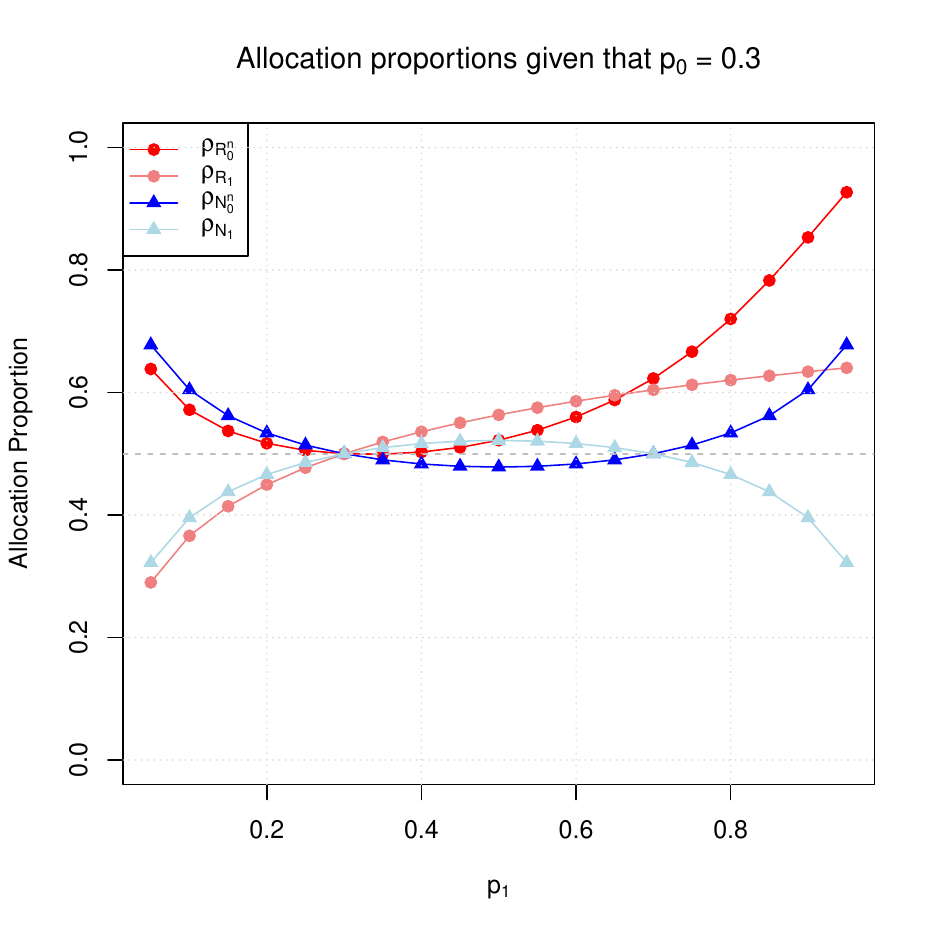}
        %\caption{Figure 1}
        \label{fig:fig1}
    \end{subfigure}
    \hfill % Space between figures
    % Second figure
    %\begin{subfigure}[t]{0.32\textwidth}
    \begin{subfigure}[t]{0.45\textwidth}
        \centering
        \includegraphics[width=\textwidth]{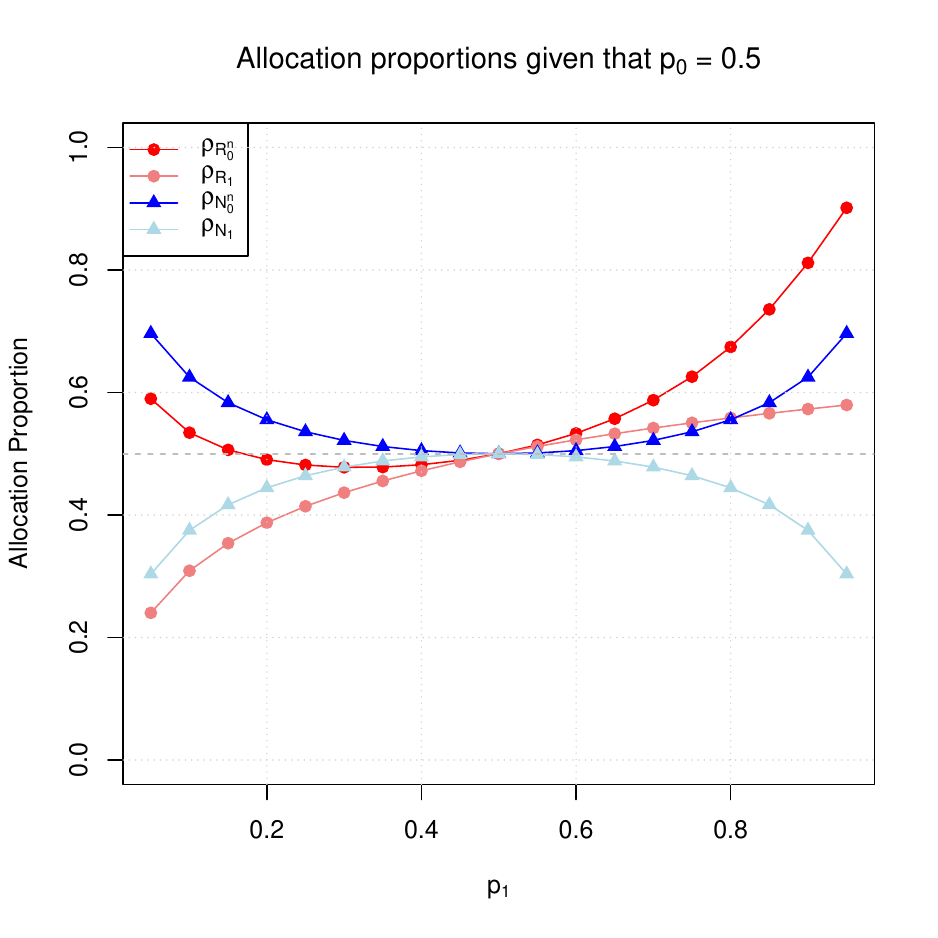}
        %\caption{Figure 2}
        \label{fig:fig2}
    \end{subfigure}
    \hfill % Space between figures
    % Third figure
       %\begin{subfigure}[t]{0.32\textwidth}
    \begin{subfigure}[t]{0.45\textwidth}
        \centering
        \includegraphics[width=\textwidth]{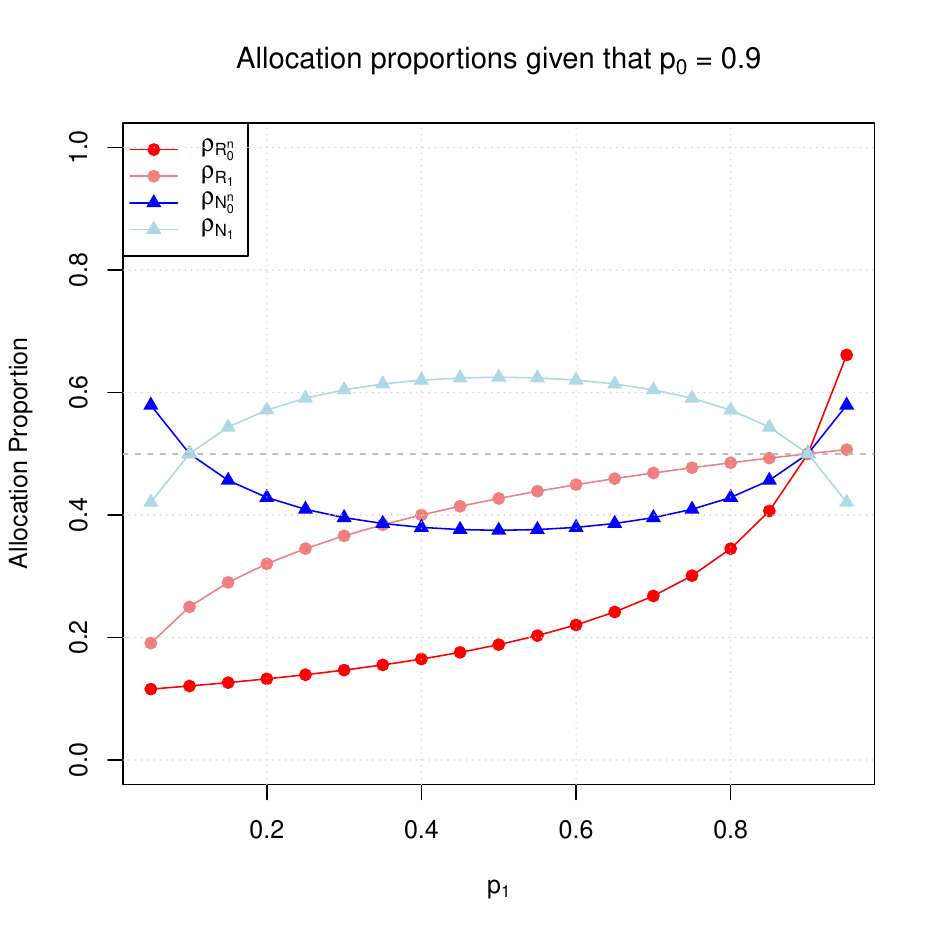}
        %\caption{Figure 3}
        \label{fig:fig3}
    \end{subfigure}
    \caption{Optimal proportions $\rho^{n}_{R_0}$, $\rho_{R_1}$, $\rho^{n}_{N_0}$, and $\rho_{N_1}$ given $p_0=0.3,0.5,0.9$ and $p_1 \in (0,1)$.}
    \label{fig:AllProportions}
\end{figure}
Figure \ref{fig:AllProportions} illustrates how the proportions change as $p_1$ varies, with $p_0$ fixed at 0.2. As indicated by the formulas, we observe that $\rho_{N_0}$ mirrors $\rho_{N_1}$. Notably, the curves for $\rho_{R_0}$ and $\rho_{R_1}$ suggest that patient allocation strategies aimed at optimizing patient outcomes differ significantly depending on the definition of the optimization problem.
These differences imply that the regions where patient-benefit and statistical power can be improved vary with the choice of the test statistic as well. Specifically, when the curves for $\rho_{R_0}$ and $\rho_{N_0}$ are sufficiently close, there is potential to both increase power and allocate more patients to the superior treatment arm. It is important to note that these regions shift based on the value of $p_0$.

\subsection{Simulation Results}

Building on the theoretical results, we now analyze the simulation for the new allocation proportions. Figure~\ref{fig:TypeIFigures} shows that the score test generally yields a lower type-I error rate than the Wald test, as seen in the comparison of CR for both tests ($\rho_{CR_0}$ vs.\ $\rho_{CR_1}$). The optimal optimal proportion $\rho^{n}_{N_0}$ successfully controls the type-I error rate and even outperform CR in this regard while the optimal proportion $\rho^{n}_{R_0}$ shows similar (or even better) performance for majority of the parametric space but is slightly inflated for values around $0.8$.

%The two optimal optimal proportions $\rho^{n}_{N_0}$ and $\rho^{n}_{R_0}$, successfully control the type-I error rate and even outperform CR in this regard.

Table~\ref{tab:1_for_ZO_Simulation} shows that \( \rho^{n}_{N_0} \) and \( \rho^{n}_{R_0} \) achieve similar power to CR, with differences within few percent in either direction across parameter combinations. 
The difference in type-I error rate control varies. The Neyman-like optimal allocation \( \rho^{n}_{N_0} \) generally provides better control of the type-I error rate compared to CR, except for values near the boundaries (i.e., close to 0 or 1). The RSHIR-like allocation proportion \( \rho^{n}_{R_0} \) shows even better type-I error rate control for the majority of the parametric space, but the type-I error rate is inflated earlier for values approaching 1.
%However, the difference in type-I error rate control is substantial: CR fails to control the nominal level, while both new proportions do. This does demonstrate that neither of the two designs using the new allocation proportion requires any of the potential solutions discussed in Section \ref{sec:Corrections}, including larger burn-in sizes.
While \( \rho^{n}_{N_0} \) may reduce the ENS compared to CR, \( \rho^{n}_{R_0} \) tends to increase it and typically allocates more patients to the superior arm.

\section{Guidance from Redesigning Trial Examples}\label{sec:TrialExample}

\subsection{Early Phase Trial}

We redesign a study to evaluate the statistical properties of an early-phase, two-armed trial designed to assess the efficacy of a treatment compared to a control. Our study was based on survival outcomes observed in a multicenter randomized phase 2 trial investigating the impact of neoadjuvant chemotherapy (NAC) in patients with locally advanced esophageal cancer (EC). The trial compared patients who received two courses of NAC versus those who received three courses, revealing that a tumor reduction rate lower than 10\% during the third course was associated with unfavorable overall survival (OS). Specifically, we use the 2-year OS rate of 89.3\% in the more favorable group and 63.5\% in the less favorable group as the basis for our assumed treatment and control arm success rates, respectively \cite{kubo2023neoadjuvant}. 
The trial was designed with a total sample size of 68 patients.

Results are summarized in Table \ref{tab:Trails}. Among the allocation proportions examined, the RSHIR-like allocation $\rho^{n}_{R_0}$, the Neyman-like allocation $\rho^{n}_{N_0}$, and complete randomization, when testing with $Z_0$, successfully control the type-I error rate. Complete randomization when testing with the Wald test exhibited slight inflation of type-I error. 
%The Neyman allocation, \(\rho^{n}_{N_0}\), for the score test reduces type-I error rate inflation compared to complete randomization, though some inflation remains. 
However, the original Neyman $\rho_{N_1}$ and RSHIR $\rho_{R_1}$ allocations resulted in highly inflated type-I error rates, making them impractical for use.  

Our findings suggest that both the RSHIR-like and Neyman-like allocation are viable methods, as they properly control the type-I error rate. While they have slightly lower power (less thatn $1\%$ difference) than complete randomization, they provide a significant increase in the expected number of successes, making them preferable approaches in this context.

\begin{table}[!htbp]
  \centering
  \begin{tabular}{llllllll}
    \hline
    \multirow{1}{*}{Test} & \multirow{1}{*}{Procedure} & Type-I error & Power & \% sup Arm (Var) & ENS \\
    \hline
    \multicolumn{6}{c}{NAC: $N=68$, $p_0=0.635$, $p_1=0.893$} \\
    \hline
    $Z_1$ & $\rho_{CR}$ & 5.5\% & 75.8\% & 0.5 (0) & 51.9 \\
    $Z_0$ & $\rho_{CR}$ & \textbf{4.7\%} & 74.2\% & 0.5 (0) & 51.9 \\
    $Z_1$ & $\rho_{N_1}$ & 65.7\% & 94.2\% & 0.2216 (0.1024) & 47.1 \\
    $Z_1$ & $\rho_{R_1}$ & 23.0\% & 76.6\% & 0.5798 (0.0207) & 53.4 \\
    $Z_0$ & $\rho^{n}_{N_0}$ & \textbf{4.6\%} & 73.6\% & 0.6064 (0.0033) & 53.8 \\
    $Z_0$ & $\rho^{n}_{R_0}$ & \textbf{4.9\%} & 73.4\% & 0.6909 (0.0076) & 55.3 \\
    \hline
    \multicolumn{6}{c}{CALISTO: $N=1502$, $p_0=0.941$, $p_1=0.991$} \\
    \hline
    $Z_1$ & $\rho_{CR}$ & 5\% & 100\% & 0.5 (0) & 1450.9 \\
    $Z_0$ & $\rho_{CR}$ & 5\% & 100\% & 0.5 (0) & 1450.9 \\
    $Z_1$ & $\rho_{N_1}$ & 96.3\% & 99.9\% & 0.1329 (0.1042) & 1423.4 \\
    $Z_1$ & $\rho_{R_1}$ & 5.4\% & 100\% & 0.5073 (0.0005) & 1451.5 \\
    $Z_0$ & $\rho^{n}_{N_0}$ & 5.2\% & 100\% & 0.7139 (0.0014) & 1467 \\
    $Z_0$ & $\rho^{n}_{R_0}$ & 5.1\% & 100\% & 0.8298 (0.0031) & 1475.7 \\
    \hline
  \end{tabular}
  \caption{Results for two different sample sizes. The first set corresponds to $N=68$, $p_{0}=0.635$, $p_{1}=0.893$ with a number of simulations $10^4$ and minimal burn-in of $2$ patients per arm. The second set corresponds to $N=1502$, $p_{0}=0.941$, $p_{1}=0.991$ with the same simulation settings.}
  \label{tab:Trails}
\end{table}

\subsection{Confirmatory Trial: CALISTO}\label{sec:CALISTO}

In this section, we redesign a published clinical trial to demonstrate the effect of the newly proposed proportions. 
We examine the CALISTO trial \citep{Decousus2010}. The goal of this trial was to assess the efficacy of a new drug, Arixtra, compared to a placebo in patients with acute symptomatic lower limb thrombophlebitis. Success was defined as the absence of a composite set of events, including death from any cause, symptomatic pulmonary embolism, symptomatic deep-vein thrombosis, symptomatic extension to the saphenofemoral junction, and symptomatic recurrence of superficial-vein thrombosis by day 47. 

The trial results revealed a success rate of 94.1\% in the placebo group and 99.1\% in the treatment group. Originally, the study was designed to include 3002 patients, but only $n=1502$ patients were included in the analysis. 
%We conducted a simulation study based on these success rates, but using a sample size of only $n=366$ patients because the original sample study was overpowered. 
We compare optimal proportions aimed at minimizing failures $\rho_{R_1}$ and $\rho^{n}_{R_0}$, as well as maximizing power, $\rho_{N_1}$ and $\rho^{n}_{N_0}$ to CR. For each method we use the test statistic $Z_0$ or $Z_1$ that was used in the optimization problem but for CR we present both.  
%The analysis involved comparing test statistics $Z_0$ and $Z_1$ for each method. 
For each method, a burn-in period of $2$ patients per arm was implemented. 
%\begin{table}[!htbp]
%  \centering
%  \caption{$N=1052$, $p_{0}=0.941$, $p_{1}=0.991$, Number of simulations $10^4$, minimal burn-in of $2$ patients per arm.}
%  \label{tab:CALSITO1502}

%  \begin{tabular}{llllllll}
%    \hline
 %   \multirow{2}{*}{Procedure} & \multicolumn{2}{c}{Type-I error} & \multicolumn{2}{c}{Power} & %\multirow{2}{*}{\% sup Arm (Var)} & \multirow{2}{*}{ENS} \\
%    \cline{2-5}
  %                             & $Z_1$ & $Z_0$ & $Z_1$ & $Z_0$ & & \\
 %   \hline
   % $\rho_{CR}$ & 5\% & 5.1\% & 100\% & 100\% & 0.5 (0) & 1450.9  \\
  %  $\rho_{N_1}$ & 95.9\% & 95.9\% & 99.9\% & 99.9\% & 0.1329 (0.1042) & 1423.4 \\
 %   $\rho_{R_1}$ & 5.3\% & 5.3\% & 100\% & 100\% & 0.5073 (0.0005) & 1451.5 \\
%    $\rho^{n}_{N_0}$ & 4.6\% & 4.6\% & 100\% & 100\% & 0.7139 (0.0014) & 1467 \\
 %   $\rho^{n}_{R_0}$ & 4.7\% & 4.7\% & 99.9\% & 99.9\% & 0.8298 (0.0031) & 1475.7 \\
   % \hline
  %\end{tabular}
%\end{table}
Table \ref{tab:Trails} demonstrates that CR does lead to asymptotic type-I error control using either the Wald or score test. The proportions $\rho^{n}_{R_0}$ and $\rho^{n}_{N_0}$ almost control type-I error rate, while $\rho_{R_1}$ and more so $\rho_{N_1}$ do not.
While all designs have similar power, both proportions $\rho^{n}_{R_0}$ and $\rho^{n}_{N_0}$ are able to increase patient benefit significantly. Especially, using $\rho^{n}_{R_0}$ leads to 25 additional success compared to CR.

\section{Discussion and Future Research}\label{sec:discuss}
In this paper, we introduce two optimal allocation proportions that demonstrate type-I error rate control in adaptive two-armed trials with binary outcomes. In finite samples they achieve type-I error rate control for the largest margin of the parametric space (90\%).
%and offer the best type-I error rate control for the remaining 10\%.
%In this paper, we introduce two optimal allocation proportions that achieve type-I error rate control in adaptive two-armed trials with binary outcomes. 
These proportions offer a promising balance between statistical rigor and practical relevance. Our proposed methods demonstrate power performance that is comparable to existing approaches, with occasional gains, particularly in small sample settings. Importantly, the proposed proportions have the advantage of increasing the likelihood that patients receive the superior treatment.  

A potential concern with using estimated quantities rather than true values in the optimization process could be the risk of introducing additional variability. However, as we show in Section \ref{sec:AlternativeOptimalRAR}, this concern is unfounded; there is no meaningful increase in variability when using estimators in our proposed framework. This finding strengthens the practical applicability of our method by demonstrating that it maintains robust statistical properties even when applied in real-world scenarios where true parameters are unknown.  

Despite the promising results, some limitations remain. We focused on two-armed trials with binary endpoints, as these are relevant in practice \citep{Pin2025software} and are among the most frequently discussed settings in the theoretical literature \citep{Hu2006}.
%, Pin_Deming2024}. 
\cite{azriel2012optimal} discusses the relationship to other optimal allocations for discrete and continuous endpoints that optimize for power. However, the scope of our results do not yet extend to multi-armed trials, which remain an important area for future research. 
%We would refer the interested reader to our book chapter for more details \citep{Pin_Deming2024}.

%The multi-armed case poses additional challenges \citep{tymofyeyev2007optimal, sverdlov2020optimal}, particularly in defining optimal allocation proportions, and requires further exploration to close this gap. 
While our focus in this paper is on two-arm trials, there is a large literature on optimal allocation proportions in multi-armed settings. However, the multi-armed case poses additional challenges \citep{tymofyeyev2007optimal, sverdlov2020optimal}. For instance, multi-armed extensions of Neyman allocation exist \citep{tymofyeyev2007optimal}, though they tend to allocate patients only to the best and worst arms, which is undesirable in practice. As shown in \citet{Zhu2009, Antognini2021}, additional constraints (e.g., minimum allocation proportions) are needed to ensure all arms are represented. Other authors have proposed allocations tailored to specific hypothesis tests or altered objectives \citep{Jeon2010, Azriel2014, Biswas2011}.
Our proposed proportions, which are optimized specifically for the score test in the two-arm setting, likely extend in the same way the multi-arm case but also have the same limitations. While a score-test analog of the multi-armed Neyman allocation could, in principle, be derived, it would also only allocate to two arms unless further constraints are imposed. 
\cite{tymofyeyev2007optimal} investigated multi-armed extensions of both Neyman and RSHIR allocation, but the latter solution requires the choice of a smoothing kernel function and the choice of unknown parameters, but do not provide a closed-form solution.

%Moreover, generalizations of RSIHR allocation are only available for special cases and a limited number of arms.
%We see the development of optimal allocations for multi-armed response-adaptive trials as an important direction for future work. 
%This paper aims to lay the theoretical groundwork in addition to the work by \citet{Rosenberger2001} for such developments, with the goal of extending these principles to multi-armed settings in subsequent research.

An alternative route of approaching type-I error rate control for optimal proportions would be to incorporate it directly as a constraint in the optimization problem. This approach has been discussed, for example in  \cite{baas2024constrainedmarkovdecisionprocesses}, but this would require separate computation for each sample size $n$ over the entire parameter space. Instead, we aim to highlight the practical benefits of using better suited test statistics within standard adaptive design frameworks.

Lastly, binary endpoints, such as success or failure, are a special case in that their variance depends directly on the mean, i.e., the success rate. This influences the behavior of optimal allocation rules and differentiates binary trials from most trials with continuous outcomes. While our work addresses this specific case, optimal proportions for other endpoints will look different and do not have the same properties. \cite{Ye2024} explored the robustness of RAR and
the use of the Wald test for continuous responses and large samples.

There are several promising directions for future work. First, we want to adapt our approach to alternative measures of interest, such as relative risk or odds ratios. The calculations are straightforward for the Wald test \citep{Rosenberger2001, Pin_Deming2024} and can be equivalently derived for the score test. However, these measures are widely used in practice and present different statistical properties that may influence optimal allocation strategies. Second, we aim to explore optimal allocation proportions for nonparametric tests and exact tests, which are particularly relevant in small-sample trials where traditional asymptotic methods may not perform well. Additionally, one could investigate deriving optimal proportions for finite sample corrections, such as the Agresti and Caffo correction, i.e. using the adjusted estimators in the optimization problem. 

%By integrating our expertise in response-adaptive randomization theory with our experience in clinical trials, 
This work represents a significant step towards broader implementation of these methods. It addresses the current regulatory landscape, which prioritizes both patient-centered trials and robust type-I error rate control, by providing a framework that achieves both.

% 

%  The \backmatter command formats the subsequent headings so that they
%  are in the journal style.  Please keep this command in your document
%  in this position, right after the final section of the main part of 
%  the paper and right before the Acknowledgements, Supplementary Materials,
%  and References sections. 

\section*{Acknowledgments}

The authors acknowledge the use of large language models (LLMs) to assist with generating figures and refining grammar and wording in this paper. The LLMs were not used for data analysis, interpretation, or original scientific writing. All content has been carefully reviewed and verified by the authors, who take full responsibility for the integrity and accuracy of the work.

The authors acknowledge funding and support from the UK Medical Research Council (grants MC UU 00002/15 and MC UU 00040/03 (SSV), as well as an MRC Biostatistics Unit Core Studentship (LP) and the Cusanuswerk e.V. (LP). SSV is part of PhaseV's advisory board. Additionally, LP's visit to the Department of Statistics at George Mason University to collaborate with WR on this project was supported by the Trials Methodology Research Partnership Doctoral Training Program and the Fisher Memorial Trust. LP is thankful for the departments hospitality. 
We thank Dr. Stef Baas for revising our codes and spotting two minor errors.
The authors are grateful to the AE and referees for their insightful suggestions.

\section*{Supplementary Material (Code)}

The code referenced in Section \ref{sec:newRSHIR} is available with this paper at the \textit{Biometrics} website on Oxford Academic. The code can be found at the GitHub repository \url{https://github.com/lukaspinpin/Optimal-Proportions}, and contains the R code necessary to replicate the results presented in the paper.

\section*{Data Availability}

There are no data presented in this paper.

\bibliographystyle{abbrvnat} % Alternative: unsrtnat
\bibliography{references}

 %%% Uncomment this line and comment out the ``thebibliography'' section below to use the external .bib file (using bibtex) .





% \end{thebibliography}

\end{document}